\documentclass[prb,12pt]{revtex4}
    \begin{document}
    \title{First-principles study of the origin of alkali promotion of reactivity of
     metal surfaces.}
    \author{Sergey Stolbov, Talat S. Rahman}
    \affiliation{Department of Physics, Cardwell Hall, Kansas State
    University, Manhattan, KS 66506}

    \begin{abstract}
    Alkali metals are known as efficient promoters in heterogeneous
    catalysis but the nature of this phenomenon is as yet not
    understood. Our first principles calculations reveal a huge
    increase and delocalization of the isoelectronic reactivity index
    of Cu and Pd surfaces on alkali adsorption. We trace this
    phenomenon to an unusual feature in the surface potential formed
    by the adsorbate and propose it as a driving force for the
    "promotion effect". Our results also provide clues to some of the
    unusual optical properties of quantum wells formed when alkali
    adsorb on Cu surfaces.
    \end{abstract}

    \maketitle Adsorption of alkali metals on transition or noble
    metal surfaces dramatically changes properties of the latter
    making these systems the subject of extensive studies in areas
    extending from fundamental physics to chemical engineering. First,
    alkalis work as efficient promoters in heterogeneous catalysis, as
    their small amount adsorbed on a catalyst surface substantially
    enhances the latter's reactivity. The global character of the
    promotion effect of alkali metal is also impressive. The
    enhancement of reactivity under alkali adsorption has been
    reported for such varied catalysts as noble metals (Ag, Cu),
    transition metals with varying number of valence $d$-electrons
    (Fe, Co, Ni, Pt, Pd, Ta, Rh, Ru) and, oxides like FeO
    \cite{ert79,whi88,cup02,maa03}. The effect extends to various
    types of reactions, such as dissociation of O$_2$, NO, N$_2$ and
    CH$_4$, oxidation of CO, and methanol synthesis
    \cite{ert79,whi88,cup02}. Comprehending the nature of the change
    produced by alkali adsorption is thus essential for facilitating
    the design of new and efficient catalysts. In addition to the
    effect on surface reactivity, the formation of quantum wells
    has been observed on Na and Cs adsorption on
    Cu(111) \cite{lin87,car00,chu03}. For the Cs/Cu(111) system,
    extremely long-lived excited states have also been reported
    \cite{oga99,bor01}. An understanding of these phenomena
    is important because of their impact on processes such as
    vibrational excitations, surface scattering, and
    photochemical reactions. The quantum-well states may well modify
    surface reactivity which brings us back to issues in catalysis.

    Since catalytic processes are driven by the modification of
    chemical bonds and electronic charge transfer, understanding
    of their enhancement naturally entails examination of the
    electronic structure of catalyst surfaces. An important clue
    from experimental finding is that
    alkali adsorption on metal surface causes a significant reduction
    of its work function, as reported for Na and Cs adsorbed on
    Ru(0001) \cite{pir91}, K on Pt(111) \cite{kis83}, K on Rh(100),
    and Na on Cu(110) \cite{leh96}. To explain the correlation between
    the promotion and reduction of the work function, it has been
    suggested that by bringing the Fermi level ($E_F$) closer to the
    vacuum energy, alkali metals facilitate charge transfer between
    the reactant and the catalyst surface, thereby enhancing
    reactivity \cite{ert79,bon84}. There are also assumptions that
    alkali promote reactivity through depopulation of surface states
    \cite{ber95} and that reduction of work function upon alkali
    adsorption causes "some long-range perturbation" to the electronic
    structure, which strongly affects surface reactivity \cite{bro91}.
    Another approach focuses on the electrostatic interaction between
    the reactants and the adsorbed alkalis \cite{lan85}. In a more recent
    development \cite{mor98}, the proposed rationale is that the adsorbed
    alkalis deplete the electric field caused by the surface barrier,
    leading to a reduced interaction of the reactant dipole moments
    with this field and subsequent enhancement of their reactivity. There
    are thus indications that alkalis promote surface reactivity either through
    change in the surface potential itself, or through modification of
    the electronic states around the surface induced by that change in
    the potential. While these intuitive arguments have been very helpful
        in focusing the discussion on the subject, they have stopped short of
        establishing the microscopic mechanisms responsible for reactivity
        enhancement.

    There are also other models of surface reactivity that, to our
    knowledge, have not yet been applied  to the issue of alkali adsorption
    problem. Some of them correlate surface reactivity to the local
    densities of electronic states (LDOS) of surface atoms, more
    specifically, to LDOS at the Fermi-level ($N(E_F)$) \cite{fei84}
    or to the energy of the center of the local $d$-band \cite{ham96}.
    Others originate from the theory of chemical reactivity of
    molecules \cite{par89} that characterizes the reactivity by
    the local softness function $s(r)$ \cite{pea66} which is a measure
    of the electronic response to the change in the number of electrons
    in the system. In the context of metal surfaces $s(r)$ can be
    regarded as the electronic response to $N(E_F)$ \cite{wilprb}.
    Related to $s(r)$, but more easily tractable from first principles
    calculations, is the isoelectronic reactivity index $w^N(r)$
    \cite{wil96} written as
    \begin{equation}
    w^N(r)=\frac{1}{k^2T_{el}}\left[
    \frac{\partial{\rho(r,T_{el})}}{\partial{T_{el}}} \right]
    _{V(r),N}\approx \frac{\rho(r,T_{el})-\rho(r,0)}{(kT_{el})^2}
    \end{equation}
    Here $\rho$ is the electronic density, $V(r)$ denotes an external
    potential, $N$ is the number of electrons in the system, $k$ is the
    Boltzmann's constant and $T_{el}$ is the electronic
    temperature. An increase in $T_{el}$ from "0" to a finite value creates
    electron-hole pairs with the Fermi-Dirac distribution.
    These excitations cause an electronic density variation which
    is reflected in the last term of Eq. (1). If we treat the changes
    in  $T_{el}$ as a uniform perturbation, $w^N(r)$ is then the
    electronic density response to such a perturbation. To establish
    the validity of $w^N(r)$ as a measure of reactivity, it was
    shown \cite{wil96} that for Pd(100) the spatial distribution
    of $w^N(r)$ allows rationalization of the preferred dissociation
    path for H$_2$ and the preferred chemisorption sites for H
    adatoms, in good agreement with the results of first principles
    calculations \cite{wilprb}.

    In the present paper, we show for the first time that
    alkali adsorption leads to a giant increase and
    delocalization of $w^N(r)$ towards the vacuum.  We further show
    that this phenomenon is caused by an unusual feature in the
    surface potential formed by the adsorbates which we propose to
    be the driving force for the alkali "promotion effect".

    Our calculations are based on density functional theory in the local
    density approximation. To attest to the generality of the
    phenomenon, we present results for two prototype metal surfaces
    chosen because of their obvious differences in $d$-band fillings.
    To calculate the electronic structure, surface potential and
    $w^N(r)$ of clean and adsorbate covered Pd(111) and Cu(111) we
    use the full-potential linearized augmented plane wave (FLAPW) method
    \cite{sin94} as embodied in the Wien2k code \cite{bla01}. We mimic
    the surface system with a supercell consisting of 8 atomic layers
    of Cu or Pd and 11 \AA \/ thick vacuum. In the case of systems
    with adsorbates, we use a two-dimensional (2x2) unit cell
    consisting of four metal atoms per layer plus one alkali at
    the three-fold fcc hollow site on each side of the slab,
    corresponding to 0.25 monolayer coverage. A fragment of the
    surface with the adsorbate is shown in Fig. 1. Previous calculations
    have shown this to be the preferred adsorption site. Good
    convergence in the calculated total energy is obtained by
    sampling 25 k-point mesh in the two dimensional surface Brillouin
    zone and the basis cutoff at $RK_{max}=7$. The exchange-correlation
    part of the potential is calculated in the local density
    approximation within the density functional theory \cite{per92}.
    The $w^N(r)$ function is calculated for $T_{el}=0.01 Ry$.

    Our calculations indicate that LDOS of the surface Pd atoms
    are practically not changed upon alkali adsorption.  Within
    the FLAPW method LDOS represent electronic states located
    inside muffin-tin (MT) spheres. These are mostly $d$-states
    in the cases of Pd and Cu. We can thus conclude
    that the alkali adsorption does not affect the localized $d$-state
    of the metal surfaces. In contrast to the effect on the
    LDOS, the impact of K on the isoelectronic reactivity of the
    surfaces is found to be dramatic. Fig. 2 shows $w^N(r)$
    plotted for Pd(111) and K$_{0.25}$/Pd(111) along the plane
    perpendicular to the surface. The projection of this plane is
    shown in Fig. 1. Adsorption of potassium induces a huge increase
    in $w^N(r)$ and its delocalization towards vacuum. Interestingly,
    the effect is present not only around the adsorbate atom, but also
    for the entire surface area. A similar and even larger effect is
    found for a 0.25 monolayer of Na on Cu(111). The effect can be
    illustrated more quantitatively in the one-dimensional plots
    of $w^N(r)$. The $w^N(r)$ for both systems, plotted along the
    surface normal passing fcc hollow site, which is maximally distant
    from the adsorbate atoms, is presented in Fig. 3. On the clean
    metal surface $w^N(r)$ decays rapidly, whereas, upon alkali
    adsorption, it is both enhanced and delocalized over 2 -- 4 \AA \/
    away from the surface. This remarkable behavior of
    $w^N(r)$ implies that alkali adsorption dramatically increases the
    density of low-energetic excited states in the vacuum region in
    the vicinity of the surface facilitating an increase in surface
    reactivity.

    For insights into the microscopic basis for these dramatic effects
    we have calculated the total self consistent potential for the
    surfaces under consideration. One-dimensional plots of the
    potentials in Fig. 4 testify to a reduction of the potential
    barrier at the surface which further manifest itself in the
    well-known decrease in the work functions for Pd(111) and Cu(111).
    What is indeed very intriguing and novel is that instead of an
    expected simple reduction of the barrier, alkali adsorbates also
    form a groove-like or plateau-like region of further reduced
    potential in the vicinity of the surfaces. Since this feature is
    energetically located around $E_F$, it leads to delocalization of
    the low energy electronic excitations. Bearing in mind that
    $w^N(r)$ characterizes the density of these excitations, we can
    conclude that the striking features in the surface potential are
    responsible for the characteristics found here for $w^N(r)$ which
    in turn is a measure of the surface reactivity. It is important to note that
    we find the increase in $w^N(r)$ for metals as varied as Cu with
    almost completely occupied $d$-band and low $N(E_F)$, and for Pd
    which has partially empty $d$-band and very high $N(E_F)$. It is
    thus reasonable to expect that the phenomenon is a general
    characteristic for most metals. This is in agreement with the
    observation that alkalis promote reactions on metals with
    differing number of $d$-electrons, as mentioned earlier.

    The novel features in the surface electronic structure found here,
    may also have other consequences. For instance, the delocalization
    of $w^N(r)$ creates an unusual condition in which entire reactant
    molecule (like CO) may be "wrapped" with low-energy electronic
    excitations. Thus, the surface may affect the reactant not only
    through direct surface-reactant chemical bonds, but also through
    modification of the character of intra-molecule bonds caused by
    these excitations. Increasing the electronic response, they may
    also lead to vibrational softening.  Since the excitations are
    substantially extended toward vacuum, they should affect the
    behavior of approaching reactant well before it forms a chemical
    bond with the surface atoms. The character of the reactant-surface
    interaction may thus be modified at the early stages of adsorption
    bringing about changes in the physisorption energetics.  In the
    same spirit, the groove-like or plateau-like features of the
    potential revealed in our calculations create wide regions in the
    vicinity of the surface with almost zero electric field which may
    provide the rationale for the promotion mechanism proposed in Ref.
    \cite{mor98} (see above).

    Our findings are also relevant to the unusually long-lived excited
    states observed for Cu(111) with adsorbed alkalis
    \cite{chu03,oga99,bau02} which have been proposed to be the result
    of the strong polarization of the anti-bonding adsorbate states
    away from the surface, thereby reducing the probability of
    transition between them and the quantum well surface states
    \cite{bor01,bau02}. The potential well outside the surface found
    in our calculations could be the basis for this polarization.
    Interestingly, single-electron calculations overestimate lifetime
    of these excited states and good agreement between the calculated
    and observed lifetime is obtained only if a strong
    electron-electron interaction is assumed \cite{chu03}. It has been
    suggested that this electron-electron interaction is the result of
    enhanced electronic screening outside of the surface. Since
    $w^N(r)$ characterizes the electronic response (electronic
    susceptibility of the system), its high value and delocalization found in our
    work is evidence that the screening outside the surface is indeed
    strong.

    In summary, our calculations reveal a giant increase and
    delocalization of the isoelectronic reactivity index of the metal
    surfaces upon alkali adsorption which we propose to be the driving
    force for the well-known alkali "promotion effect". We find that
    the phenomenon is accompanied by a groove-like or plateau-like
    feature in the surface potential formed on the alkali adsorption.
    Our findings also provide the rationale for the unusually
    long-lived excited states observed for Cu(111) with adsorbed
    alkalis.  The phenomenon is also expected to be a general
    characteristic of alkali adsorption on metal surfaces, as exemplified
    in this work through the application to two very different metal
    surfaces.

    \begin{acknowledgements}
    We thank G. Ertl for helpful discussion, and acknowledge financial
    support from DOE under grant No. DE-FGO3-03ER15464.
    \end{acknowledgements}

    \section{Figure Captions}

   Fig. 1. (Color online) Top view of the A$_{0.25}$/M(111) surface (A=Na, K;
   M=Cu,Pd). The large balls represent alkali atoms,  small dark balls are
   for the first layer M atoms and small grey balls are for the rest of the M
   atoms. The dashed line represents the projection of the plane along which
   $w^N(r)$ is plotted in Fig. 3.

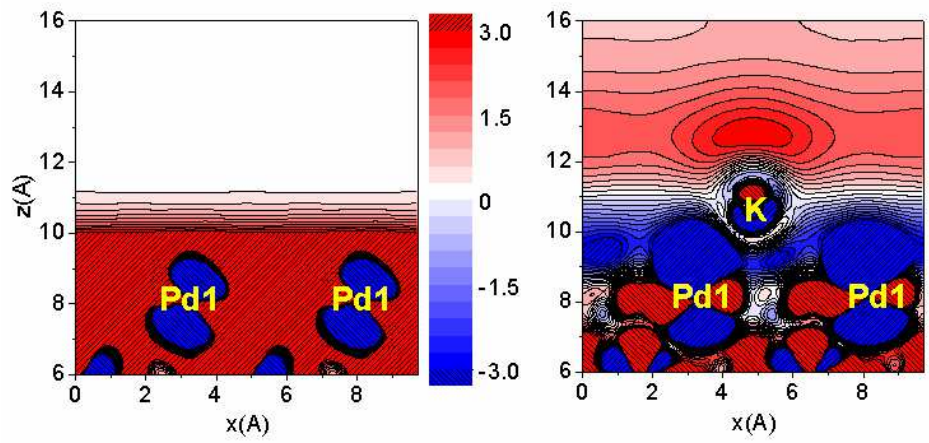
    Fig. 2. (Color online) Plot of $w^N(r)$ for clean Pd(111) (left panel) and for
    K$_{0.25}$/Pd(111) (right panel) along the plane perpendicular to
    the surface. Pd1 mark the positions of the topmost Pd atoms and K
    that of  potassium atoms. The white area corresponds to $w^N(r)=0$.

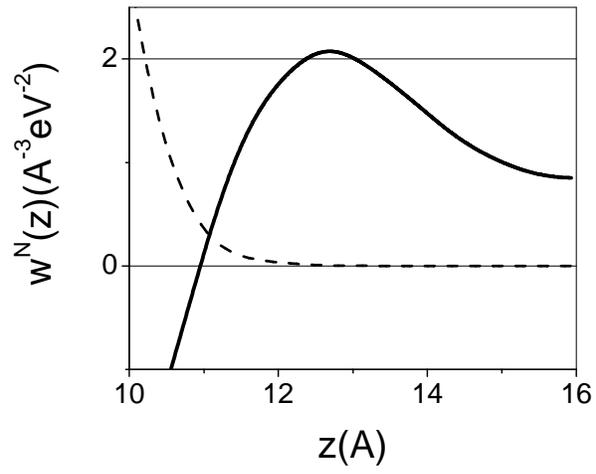
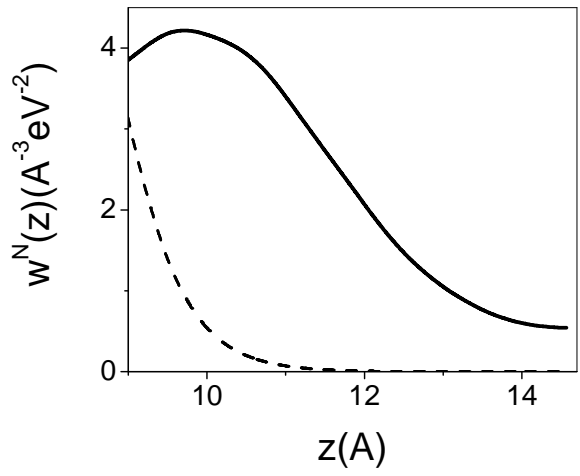
    Fig. 3. Plot of $w^N(r)$ along the surface normal. The upper panel
    shows $w^N(r)$ for Pd(111) (dashed line) and K$_{0.25}$/Pd(111)
    (solid line). The lower panel shows $w^N(r)$ for Cu(111) (dashed
    line) and Na$_{0.25}$/Cu(111) (solid line).

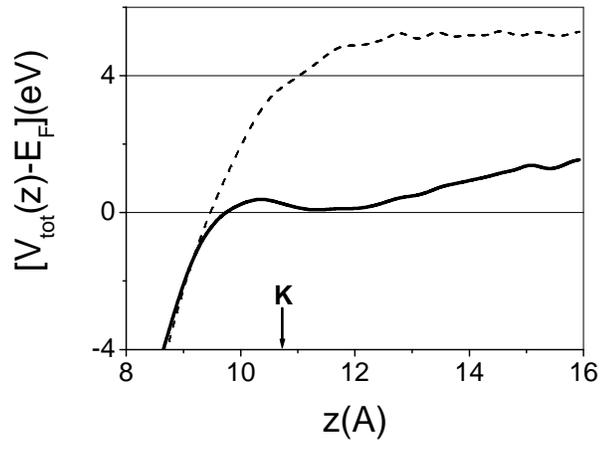
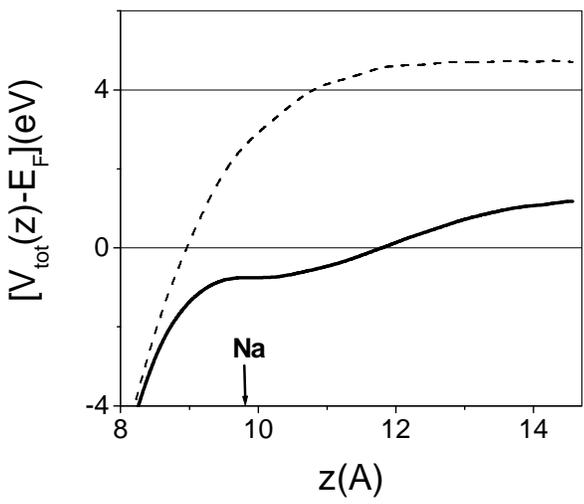
    Fig. 4. The self consistent total potential plotted along the
    surface normal. The upper panel displays the potential for Pd(111)
    (dashed line) and K$_{0.25}$/Pd(111) (solid line). The lower panel
    provides the same for Cu(111) (dashed line) and
    Na$_{0.25}$/Cu(111) (solid line). Arrows indicate the heights of
    the alkali atoms.

    \end{document}